\documentclass[aps,pra,reprint,showpacs,superscriptaddress]{revtex4-1} 
\usepackage[utf8]{inputenc}
\usepackage[fleqn]{amsmath}
\usepackage{amssymb}
\usepackage{amsmath}
\usepackage{url}
\usepackage[margin=1in]{geometry}
\usepackage{graphicx} 
\usepackage{rotating}
\usepackage{footmisc}
\DefineFNsymbols{mySymbols}{{\ensuremath\dagger}{\ensuremath\ddagger}\S\P
   *{**}{\ensuremath{\dagger\dagger}}{\ensuremath{\ddagger\ddagger}}}
\setfnsymbol{mySymbols}

\begin{document}


\title{Many Correlation Tables are Molien Sequences}

\author{Bradley Klee}
\email{bjklee@email.uark.edu, bradklee@gmail.com} 
\affiliation{Department of Physics, University Arkansas, Fayetteville, AR 72701}

\date{\today}

\begin{abstract}
Using the Jordan-Schwinger form of the quantum angular momentum eigenstates, 
it is straight-forward to define rotational correlation tables such that the 
columns are Molien sequences for finite rotational subgroup $G$. This realization gives a new and 
better means of calculation. Generalization to unitary symmetry $U(n)$ implies many more 
sequences, which determine degeneracy observables in the context of electronic, 
vibrational, and rotational motion. This leads us to discuss one physical significance 
of the Hilbert finite basis theorem. 
\end{abstract}


\maketitle 

\section{Introduction}
The use of correlation tables to capture degeneracy of eigenstates due to subgroup 
structure $G \supset SG$ dates back to the early years of the new quantum mechanics \cite{BetheGerman}. 
Under any perturbation that breaks Hamiltonian symmetry from $G$ to $SG$, elements of a 
correlation table, also called $f$-numbers, determine splitting of eigenstates. 
Level-splitting directly affects spectral measurements such as the intensity of 
dipole absorption whenever a laser irradiates a solid or molecule. Ease of accessibility 
gives the $f$-numbers importance as quantum observables. 

As an example, consider quantum angular momentum. Spherical symmetry $SO(3)$ admits circular dihedral
symmetry $D_{\infty}$ as a subgroup, so we expect to find non-trivial degeneracy in the absence of symmetry-breaking 
perturbations. The well-known eigenstates have quantum numbers $(j,m)$, which admit the 
correlation of Table I. 
\begin{center}
\textbf{Table I.} $SO(3) \supset D_{\infty}$ Correlation.
\begin{tabular}{ c || c | c | c | c | c | c | c | c}
\hline
  $j \backslash m$ &  $\;\; 0 \;\;$ & $\pm 1/2 $& $\pm 1$& $\pm 3/2 $& $\pm 2$ &$\pm 5/2$ &$\pm 3$  & \; $\cdots$ \; \\
\hline
\hline
$0$   &  $1$ & $0$ &  $0$ & $0$& $0$ & $0$ & $0$  & $\cdots$ \\
$1/2$   &  $0$ & $1$ &  $0$ & $0$& $0$ & $0$ & $0$  & $\cdots$  \\
$1$   &  $1$ & $0$ &  $1$ & $0$& $0$ & $0$ & $0$ & $\cdots$   \\
$3/2$   &  $0$ & $1$ &  $0$ & $1$& $0$ & $0$ & $0$  & $\cdots$  \\
$2$   &  $1$ & $0$ &  $1$ & $0$& $1$ & $0$ & $0$  & $\cdots$  \\
$5/2$   &  $0$ & $1$ &  $0$ & $1$& $0$ & $1$ & $0$  & $\cdots$  \\
$3$   &  $1$ & $0$ &  $1$ & $0$& $1$ & $0$ & $1$  & $\cdots$  \\
$\vdots$  & $\vdots$  & $\vdots$ & $\vdots$  & $\vdots$ & $\vdots$ & $\vdots$ & $\vdots$  & $\ddots$
\end{tabular}
\end{center}

\begin{figure}[ht]
\begin{center}
\includegraphics[scale=.4]{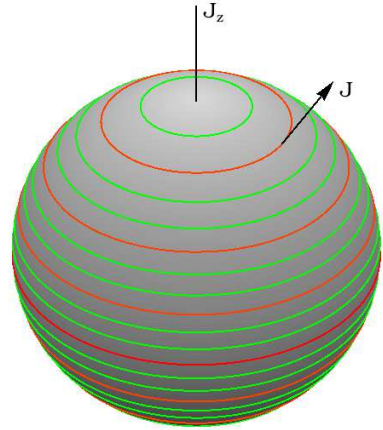} 
  \caption{Rotational Energy Surface. Weak perturbations do little to distort the shape of the RES. 
  Trajectories at $j=10$ are colored according to representations of $D_3$ symmetry: $A_0$ red, $A_0 / A_1$ red-orange, $E_1$ green. }
  \label{fig:RESPicture}
\end{center}
\end{figure} 

The spectra of a rotating molecule follows from the specification of a Hamiltonian, 
$\Omega( \mathbf{J} )$, a function of quantum mechanical angular momentum variables 
$\mathbf{J}=( J_x,J_y,J_z )$. In a semi-classical theory of rigid quantum rotations, 
the Hamiltonian determines a rotational energy surface (RES) where the $\mathbf{J}$ vector 
moves, approximately, along quantized level sets \cite{Handbook}. Description of 
planar molecules involves a descent of symmetry 
\begin{equation}
\Omega =   \frac{\omega}{\hbar^2} \; \mathbf{J} \cdot \mathbf{J} + 
\frac{\xi_z}{\hbar^2} \;\mathbf{J} \cdot \mathbb{P}_z \cdot \mathbf{J}  \;  + 
\xi_{\theta} \; \theta(\mathbf{J} / \hbar ),
\end{equation}
where $\mathbb{P}_z $ is a uniaxial projection matrix and $\theta$ is a higher-order 
polynomial function of the $\mathbf{J}$ variables. The approximate symmetry 
of the atomic nuclei constrain the range of possibilities for $\theta$. Fig.~\ref{fig:RESPicture}
depicts a spherical RES with slight perturbations. 

Frequency hierarchy $\xi_{\theta} \ll \xi_z \ll \omega  $ characterizes the simplest case where, to order $\xi_z$,
the RES is an ellipsoid with just one circular cross section. Frequency levels split 
according to Table I. At $j=3$, we have doublets corresponding to $m = \{ \pm 1, \pm 2, \pm 3 \}$ 
as well as one singlet $m=0$. Including terms to order $\xi_{\theta}$ lifts some, possibly all, 
of the remaining degeneracy. 

Calculations of rotational frequency levels and spectroscopic measurements of octahedral molecules 
at high-$j$ reveal intriguing degeneracy sequences \cite{HarterLetter}. Considering a great 
number of excited states, it becomes possible to extend upon and improve the symmetry analysis 
available through the original crystal field theory. Harter explains how patterns of 
degeneracy follow the $f$-numbers of various correlations between spherical, octahedral, and cyclic 
symmetry groups \cite{HarterRESI}. 

The standard theory common between solid state physics and molecular 
physics \cite{HarterBook} defines rotational correlations according to a character 
formula 
\begin{equation}
CT(SO(3) \supset G) = \widetilde{\chi} \cdot \chi^{-1}.
\end{equation}
where $\chi$ and $\widetilde{\chi}$ are matrices of characters, the traces
of matrices that represent group actions. Each row of character table $\chi$ corresponds to an 
irreducible representation (irrep) of $G$, while each column corresponds 
to a conjugacy class of $G$. The rows and columns of $\widetilde{\chi}$ 
correspond to irreps of $SO(3)$ and the conjugacy classes of $G$. The book \textit{Principles of Symmetry, 
Dynamics, and Spectroscopy} \cite{HarterBook} completely explains character theory, up to 
the derivation of rotational tables in Chapter 5. 

General and important features of rotational correlation tables are known and utilized
since 1929: The table rows have a periodic structure. Whole integer representations 
of angular momentum correlate to representations of the subgroup $G$ while half-integer 
representations correlate only to representations of the double subgroup $2G$. According 
to these observations it should be possible to write the correlation tables as
a set of generating functions.

Referencing various column sequences in the On-line Encyclopedia of Integer Sequences (OEIS) \cite{OEIS}, 
we find many suggestive connections between rotational correlation tables and various 
Molien series. Following the references through diversions into coding theory \cite{Sloane} 
and combinatorics \cite{Stanley}, we make the realization that classical invariant 
theory also applies to quantum rotations and vibrations. The following presentation 
provides an alternative, or at least a supplement, to the standards currently available.

\section{Quantum Invariant Theory}
In the history of science, the advent of quantum transformation theory follows after classical invariant theory \cite{GroupHist}. 
The old invariant theory, in its general form, is founded upon famous theorems by Hilbert, Noether,
and Molien. These century old theorems have practical value in a wide range of applications \cite{Sloane},\cite{Stanley}. 
They apply to quantum mechanics wherever symmetry breaking lowers unitary to a finite subgroup. We show that classical and 
quantum transformation theories admit a natural and fruitful combination by considering polynomials 
of commuting variables, the familiar quantum harmonic oscillator raising operators $a_i^{\dagger}$. 

The $SU(2) \sim SO(3)$ double cover requires the use of double groups. This critical insight 
allows us to improve and extend the notion of a rotational correlation table. By our definitions, 
Molien's theorem implies the following corollaries 
\begin{itemize}
 \item \textbf{Corollary 1.} The $f$-numbers of rotational correlation table $CT(U(2)\supset 2G:\Gamma_{1/2})$
 are determined entirely by the Molien equation when $\Gamma_{1/2}$ is a $j=1/2$ representation of $G$.
\end{itemize}
\begin{itemize}
 \item \textbf{Corollary 2.} For any $n$-dimensional irreducible representation $\Gamma$ of finite 
 symmetry group $G$, the $f$-numbers of correlation table $CT( U(n) \supset G:\Gamma )$ are 
 determined entirely by the Molien equation. 
\end{itemize}

In stating these corollaries we use a naming convention slightly different from Eq. 2, which is required 
in a wider context that applies throughout the Born-Oppenheimer hierarchy; to the electronic, vibrational, 
and rotational motion of molecules and solids. 

\subsection{Molien Equation}
Molien's equation,
\begin{eqnarray}
gf(\Gamma_x,\Gamma_y,\lambda)  = \frac{1}{|G|}\sum_{A_i \in \Gamma_x} \frac{\chi^{*}(\Gamma_y,i)}{Det[\mathbb{I}-A_i \; \lambda]} \\
                               =  f_0(\Gamma_y) +f_1(\Gamma_y)\lambda +f_2(\Gamma_y)\lambda^2+\; \; ... \nonumber
\end{eqnarray}
is a clever utility that automates the analysis of induced representations, thus enabling the 
counting of invariants and covariants. 

Representation $\{ A_1,A_2,...,A_g\} \in \Gamma_x$ is a group of  $m\times m$ 
dimensional matrices $A_i$ that act linearly on polynomial variables 
$\mathbf{x} = (x_1,x_2,...,x_m)$. This one representation of $G$ determines an infinite set 
of representations $\{ A_1^n,A_2^n,...,A_g^n\} \in \Gamma_x^n$. Each representation 
$\Gamma_x^n$ acts linearly on a complete set of polynomials of homogeneous order $n$, 
$\mathbf{x}^n = (x_1^n,x_1^{n-1}x_2,..., x_2^{n},x_2^{n-1}x_3,..., x_m^n)$. 
Notice that a transformation of the $\mathbf{x}$ variables always causes an element of $\mathbf{x}^n$ to transform 
into a linear combination of all elements of $\mathbf{x}^n$. 

A subspace $\mathbf{s}^n = \mathbb{P}^n_s\mathbf{x}^n \in \mathbf{x}^n$ is said to transform covariantly to irrep
$\Gamma_y$ whenever the $\Gamma_y$ projector acts as identity
\begin{equation}
\mathbf{s}^n =  \mathbb{P}^n_{\Gamma_y} \mathbf{s}^n.
\end{equation}
If $\Gamma_y$ is the trivial irrep with all characters equal to one, then $\mathbf{s}^n$ is also said to transform 
invariantly. This definition gives the projectors $\mathbb{P}^n_{\Gamma_y}$ central importance
in any attempt to enumerate sequences of covariants.

To define the idempotent projectors \cite{HarterBook}, we need the irreducible characters $\chi(\Gamma_y,i)$ of 
the $i^{th}$ element of group $G$. These characters are listed by conjugacy class 
in the character tables. Then we have, 
\begin{equation}
 \mathbb{P}^n_{\Gamma_y} = \frac{1}{|G|}\sum_{A_i \in \Gamma_x} \chi^{*}(\Gamma_y,i) A^n_i,
\end{equation}
where the $*$ indicates complex conjugation. Every polynomial of homogeneous order $n$ transforms 
covariantly to some irrep according to the decomposition of the identity operator
\begin{equation}
 \mathbb{I}^n = \sum_{y}\chi(\Gamma_y,1)\mathbb{P}^n_{\Gamma_y},
\end{equation}
where $i=1$ is assumed to index the identity operation $\mathbb{I} = A_1$, and 
the sum is taken over all irreps. 

Clearly the task of counting the number of polynomial subspaces with some particular transformation 
property reduces to computation of projector traces
\begin{equation}
f_n(\Gamma_y) = Tr[\mathbb{P}^n_{\Gamma_y}],
\end{equation}
where by Eq. 6, $f_n(\Gamma_y)$ counts each $\chi(\Gamma_y,1)$-dimensional multiplet exactly once. 
According to properties of the trace function, $Tr[\mathbb{P}^n_{\Gamma_y}]$ is determined 
from the quantities $Tr[A^n_i]$ alone. 

Fortunately, Molien notices that 
\begin{equation}
\frac{1}{d(\Gamma_x,i,\lambda)} = \frac{1}{Det[\mathbb{I} - A_i \;\lambda ]} = \sum_n Tr[A_i^n]\lambda^n, 
\end{equation}
thus it becomes unnecessary to calculate induced representations explicitly. Finally, 
the proof of the covariant Molien theorem rests upon the same crux as the invariant 
theorem nicely re-proven by Sloane \cite{Sloane}.

\begin{itemize}
 \item \textbf{Theorem 1. (Molien)} Coefficient $f_n(\Gamma_y)$ of $\lambda^n$ in the power series expansion of
 the generating function $gf(\Gamma_x,\Gamma_y,\lambda)$ gives the number of $\Gamma_y$-covariants of homogeneous order $n$ in 
 variables $\mathbf{x}=(x_1,x_2,...,x_m)$ that transform according to a representation $\Gamma_x$ of group $G$.
\end{itemize}

We apply this theorem in a physical context. 
\subsection{The Pseudotop Analogy}

The pseudotop analogy\cite{KleeDissertation} occurs in quantum mechanics wherever a quantum system 
with fixed spatial orientation transforms according to a rotational algebra. 
Numerous resources 
\cite{HarterBook},\cite{Schwinger},\cite{IachelloBook},\cite{KleeDemonstration} 
discuss this analogy in the context of the two-dimensional isotropic quantum harmonic 
oscillator. Here the pseudotop analogy takes a most simple form, which 
is both a sincere curiosity and a useful aid to practical calculations. 
Recall the Jordan-Schwinger form for the eigenstates of angular momentum
\begin{equation}
|j,m\rangle \propto (a_1^{\dagger})^{j+m} (a_2^{\dagger})^{j-m}|0,0\rangle, 
\end{equation}
where $(j,m)=\frac{1}{2}(n_1+n_2,n_1-n_2)$ provides a relation between rotational quantum numbers 
$(j,m)$ and vibrational quantum numbers $(n_1,n_2)$. 

Polynomials of the commuting raising operators $\boldsymbol\eta=(a^{\dagger}_1,a^{\dagger}_2)$ determine all 
representations of continuous group $SU(2) \sim SO(3)$. As in section IIa we 
define induced spaces $\boldsymbol\eta^{2j}$ where polynomials of the $\boldsymbol\eta$ variables have homogeneous order $2j$. 
All polynomials transform invariantly by identity symmetry, so the trivial Molien equation in 
two-dimensions gives the generating function for the total degeneracy
\begin{eqnarray}
 gf(\{(^{1}_0\;^{0}_{1})\},\{(1)\},\lambda) = 1/(1-\lambda)^2 \\
 = 1+ 2 x + 3 x^2 + 4 x^3 + 5 x^4  ... \nonumber
\end{eqnarray}
The series coefficients equal the sum of columns in Table I above. 

In a situation where finite rotational symmetry $G$ constrains a Hamiltonian with approximate spherical 
symmetry $SU(2) \sim SO(3)$, we would also like to impose the constraints of $G$ onto the eigenstates. This is 
usually done using projectors $\mathbb{P}^n_{\Gamma_y}$, but classical invariant theory provides 
an alternative approach that we like to explore. 

Our conventions present the only difficulty in applying Theorem 1 directly. Usually we start with a 
real-space, $j =  1$ representation $\Gamma_1$, but $\boldsymbol\eta$ transforms according to the $j=1/2$ 
representation $\Gamma_{1/2}$. To perform reverse-induction, define 
\begin{equation}
\eta^2_i  =  \boldsymbol\eta \cdot g_i \cdot \boldsymbol\eta ,
\end{equation}
where $\mathbf{g}= \frac{1}{2} (\mathbb{I}+\sigma_3, \sqrt{2} \; \sigma_1, \mathbb{I}-\sigma_3)$ with $\sigma_i$ the standard 
Pauli matrices.

The matrix/vector $\mathbf{g}$ transforms according to either 
\begin{subequations}
\begin{align}
 \mathbf{g} \longrightarrow \mathbf{g}' &= \mathcal{D}^{1}_{\alpha,\beta,\gamma} \cdot \mathbf{g}, \\
 g_i \longrightarrow g_i' &= (\mathcal{D}^{1/2 }_{\alpha,\beta,\gamma})^T 
 \cdot g_i \cdot \mathcal{D}^{1/2}_{\alpha,\beta,\gamma} \;\;,
\end{align}
\end{subequations}
where $\mathcal{D}^{j}_{\alpha,\beta,\gamma}$ is a j-irrep rotation matrix with Euler angles 
$( \alpha, \beta, \gamma )$ \cite{HarterBook}. 

Equation 12 provides an equivalence between vector rotation and matrix conjugation that allows the 
determination of $\Gamma_{1/2}$ from $\Gamma_1$. A difficulty arises because the similarity 
transform allows $-\mathbb{I}$ to act as identity. This opens up the possibility that 
$\Gamma_1$ is not a faithful representation of the group $G$ represented by $\Gamma_{1/2}$, 
that $\Gamma_{1/2}$ may contain more distinct elements than $\Gamma_1$. 

The general form of $\mathcal{D}^{1/2}_{\alpha,\beta,\gamma}$ is
\begin{eqnarray}
& \mathcal{D}^{1/2}_{\alpha,\beta,\gamma} = e^{-i\frac{\alpha}{2}\sigma_3 }
e^{-i\frac{\beta}{2}\sigma_2}e^{-i\frac{ \gamma}{2}\sigma_3} \\
& =\left( \begin{array}{c c} 
 e^{-i\;(\alpha+\gamma)/2} cos \; \beta/2 & -e^{-i\;(\alpha-\gamma)/2} sin \; \beta/2   \\
 e^{i\;(\alpha-\gamma)/2} sin \; \beta/2 & e^{i\;(\alpha+\gamma)/2} cos \; \beta/2   
 \end{array} \right) \nonumber.
\end{eqnarray}
The factor of $2$ attached to all angles in Eq. 13 causes all $2\;\pi$ rotations to have a representation 
\begin{equation}
\mathcal{D}^{1/2}_{2\;\pi} = -\mathbb{I},
\end{equation}
while, intuitively, a $2 \pi$ rotation in $\Gamma_1$ must act as identity. 
This discrepancy requires that the $\Gamma_{1/2}$ representation contains twice as many elements 
as the $\Gamma_{1}$ representation and explains the locution that '$SU(2)$ is the double cover of $SO(3)$'. 
We agree with Klein \cite{Klein} that the $\Gamma_{1/2}$ representation of $G$ is actually 
a representation of the double group $2G$. 

$\Gamma_{1/2}$ takes a special place among all $2G$ irreps because of spin and the 
pseudotop analogy. By Eq. 11, $\Gamma_{1/2}$ also transforms $\boldsymbol\eta$, 
so it is exactly the representation we need to use the tools of classical invariant theory in an 
analysis of quantum rotations. Setting 
$\Gamma_x=\Gamma_{1/2}$ in the Molien equation, we have
\begin{equation}
 CT(U(2)\supset 2G:\Gamma_{1/2}) = gf(\Gamma_{1/2},\Gamma_y,\lambda),
\end{equation}
where each $\Gamma_y$ of group $2G$ determines a column of the correlation table. 
We assert that corollary 1 is proven by definition, but it is also possible to prove 
equivalence between the character formula and the Molien equation by comparing how 
each method treats the decomposition of the $Tr[\mathbb{P}^n_{\Gamma_y}]$. 
\subsubsection{Examples}
Using Eqs. 12 we generate $\Gamma_{1/2}$ from $\Gamma_1$ for each of the following groups: 
Triangular Dihedral($D_3$), Octahedral ($O$), and Icosahedral ($A_5$). From the $\Gamma_{1/2}$ representations we 
calculate correlation tables $CT(U(2)\supset 2G:\Gamma_{1/2})$ according to Molien's Eq. 2. 
These tables, available on-line \cite{KleeMolien}, are 
the same as typically obtained \cite{BetheGerman},\cite{HarterRESI},\cite{HarterIco},\cite{IcoRot} 
by character formula Eq. 2, up to an arbitrary $g$ or $u$ label. The on-line supplement  
also contains an incomplete cross-reference of the column sequences with entries in the 
OEIS \cite{OEIS}. Some of the sequences computed in this investigation seem to be missing from 
the extensive records.  

\subsection{Generalization} 
Corollary 2 generalizes corollary 1 by removing the restriction $\Gamma_x=\Gamma_{1/2}$. 
This is an important move because quantum mechanics involves many symmetries other than 
$SU(2)\sim SO(3)$. In a sufficiently general setting we have an $n$-dimensional representation 
$\Gamma_n$ of group $G$ which is also a subgroup of $U(n)$. The representation generators 
$\boldsymbol\eta_n = (a_1^{\dagger},a_2^{\dagger},...,a_n^{\dagger})$ are the ladder 
operators of the $n$-dimensional isotropic quantum harmonic oscillator.
These commuting variables we again subject to the constraints of symmetry. 

The molecular physics literature already contains numerous examples in the general 
setting, including some investigation of the Molien function\cite{Sad}. 
We review a few articles and provide Molien generating functions that 
help to explain symmetry classification of excited vibrational states. 

\subsubsection{Examples}
Level correlations in \cite{Patterson} are given by 
\begin{eqnarray}
SF_6 , \; UF_6 :  CT(U(3)\supset O:T_1), \nonumber \\
SiF_4 : CT(U(3)\supset O:T_2). \nonumber 
\end{eqnarray}
Tables available on-line \cite{KleeMolien} give $f$-numbers consistent with 
the published levels.

Level correlations in \cite{HalonenTrig},\cite{HalonenOct} 
are slightly more complicated, but nevertheless yield to a Molien analysis. 
The separate ladders obey
\begin{eqnarray}
|\nu_1,0,0\rangle &:  CT(U(1)\supset D_3:A_{1}), \nonumber \\
|0,\nu_2,0\rangle &:  CT(U(1)\supset D_3:A_{1}), \nonumber \\
|0,0,\nu_3\rangle &:  CT(U(2)\supset D_3:E); \nonumber \\
|\nu_1,0,0\rangle &:  CT(U(1)\supset O:A_{1g}), \nonumber \\
|0,\nu_2,0\rangle &:  CT(U(2)\supset O:E_{g}), \nonumber \\
|0,0,\nu_3\rangle &:  CT(U(3)\supset O:F_{1u}). \nonumber
\end{eqnarray}

Considering combintorial properties of the Molien equation it is possible 
to join separate correlation tables into one table that gives correlations 
for combined states $|v_1,v_2,v_3\rangle$. 
In these tables the $f$-numbers are coefficients of a power series expansion 
in three formal variables $\lambda_1,\lambda_2,\lambda_3$. We write the combined-ladder 
correlation tables as
\begin{eqnarray}
|\nu_1,\nu_1,\nu_3\rangle : CT(U(1)\oplus U(1)\oplus U(2)\supset \nonumber \\
D_3:A_{1}\oplus A_{2} \oplus E). \nonumber \\
|\nu_1,\nu_1,\nu_3\rangle : CT(U(1)\oplus U(2)\oplus U(3)\supset \nonumber \\
O:A_{1g}\oplus E_{g} \oplus T_{1u}). \nonumber 
\end{eqnarray}
All level degeneracies calculated using the Molien function agree with published results. 
\subsection{Finite Reduction of Eigenspaces}
The Molien equation allows us to count the number of excited states transforming covariantly 
to $\Gamma_y$ but does not exactly provide a means for enumerating the states themselves.
We could construct the projectors $\mathbb{P}_{\Gamma_y}^n$ by induction. Other authors 
\cite{HalonenOct}, \cite{IachelloLetter} call the brute force method difficult, and 
seek out new, creative approaches. According to the Hilbert finite basis theorem, 
invariant theory provides another alternative. 

Preexisting theorems ( Cf. \cite{Stanley} Thm. 1.3, 3.10 ) guarantee the 
existence of a simple procedure that generates a complete span of each $\Gamma_y$ 
eigenspace. The algorithm depends only on the input of a finite set of polynomials, which belong to 
either one of two classes \cite{Sloane},\cite{Stanley}. 
Surmounting the difficulties of troublesome syzygies, we can arrive at a 
succinct and sophisticated specification of the eigenspaces, which allows 
us to classify excited states of a perturbed oscillator. 

\subsubsection{Example}
Examining the $2D_3:E_{1/2}$ Molien functions and various idempotent projectors 
for $2j=1,2,3,4,5,6$, we determine a finite set of polynomial invariants and covariants. 
From these, we write a basis of rotational states 
\begin{equation}
|j,m,\chi\rangle \propto (a_1^{\dagger}a_2^{\dagger})^{(j-m)}((a_1^{\dagger})^{2m} +\chi (a_2^{\dagger})^{2m})|0\rangle, 
\end{equation}
with $j \geq m \geq 0$ and $\chi = \pm 1$ or $\pm i$. Apparently these are a complete, orthogonal set 
of $J_z^2$  eigenfunctions. This basis also divides nicely into classes that transform covariantly 
to the representations of $2D_3$. Table II gives the relation between quantum numbers and 
symmetry labels. 

\begin{center}
\textbf{Table II. $\Gamma_y \longleftarrow |j,m,\chi\rangle$} .
\begin{tabular}{ c || c | c | c }
\hline
\;\; $\Gamma_y$ \;\; & $\;(j-m)\;$ & $\;2m \; \% \; 6\;$ & \;\;\; $\chi$ \;\;\;  \\
\hline
\hline
$A_0$  		&  Even &  $0$ & $+1$ \\
  		&  Odd  &  $0$ & $-1$ \\  		
     		&  Even  & $m=0$  & $0$ \\ \hline  		
$A_1$   	&  Even &  $0$ & $-1$ \\
	        &  Odd  &  $0$ & $+1$ \\ 
 		&  Odd  &  $m=0$  & $0$ \\ \hline  
$E_1$   	&  $\cdot$ &  $2,4$ & $\pm 1$ \\ \hline 
$E_{3/2}^{L}$   &  Even &  $3$ & $+i$\\
		&  Odd  &  $3$ & $-i$ \\ \hline 
$E_{3/2}^{R}$   &  Even &  $3$ & $-i$ \\
	        &  Odd  &  $3$ & $+i$ \\ \hline 
$E_{1/2}$   	&  $\cdot$  &  $1,5$ & $\pm 1$ 
\end{tabular}
\end{center}

\begin{figure}[ht]
\begin{center}
\includegraphics[scale=.8]{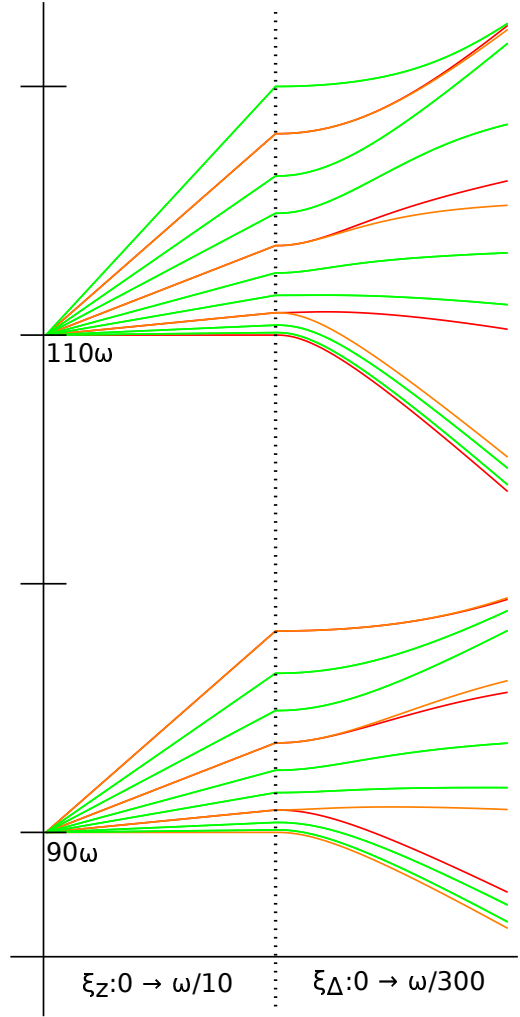} 
  \caption{Rotational Levels. The $j=9,10$ rotational levels are colored 
  according to symmetry type $A_0$ red, $A_1$ orange, $E_1$ green. From left to right, 
  a $J_z^2$ perturbation of strength $\xi_z$ is applied, followed by a $D_3$ 
  perturbation of strength $\xi_{\Delta}$. The $D_3$ perturbation causes the 
  $A_0 / A_1$ doublets to split.}
  \label{fig:FreqLevels}
\end{center}
\end{figure} 

This table is completely consistent with table $CT(U(2) \supset 2D_3:E_{1/2})$ available on-line 
\cite{KleeMolien}. We use table II and Eq. 16 to label level-splitting diagrams such 
as Fig.~\ref{fig:FreqLevels}. 

\section{Conclusion}
Direct involvement of the unitary creation operators clarifies the application 
of classical invariant theory to quantum mechanics. These polynomial variables often occur in 
molecular physics, as the generators of quantized rotations or vibrations. We give a unified 
perspective that views electronic, rotational, and vibrational multiplet counting as equivalent 
tasks. The correlation tables defined herein apply immediately to any molecule or solid
with dihedral, tetrahedral, octahedral, or icosahedral symmetry. 

As with any of the best mathematical theories, the range of applications for 
classical invariant theory is truly interdisciplinary. If we confine ourselves 
to studying physics, we still expect to find connections wherever unitary symmetry 
occurs. One as-yet unexplored possibility exists in quantum optics \cite{OpticalAngular}, 
where correlation techniques should apply to solutions of the plane-wave 
Helmholtz equation.  

This letter focuses on the simple Molien equation, which is only a small 
part of the complete invariant theory. As this research continues, we confront the 
exciting possibilities and the difficult challenges (syzygies) that arise 
in a closer examination of the Hilbert finite basis theorem. Immediately we find a 
useful equivalence between polynomial rings and irreducible eigenspaces.
Explicit construction and manipulation of creation-operator polynomials 
may lead to new discovery in physics. 

Invariant theory is an important chapter in the history of science, because it already 
contains many notions essential to the foundations of modern physics. Clebsch and Gordan 
obtained early results regarding polynomials generated by two variables. 
In the generalization to an arbitrary number of polynomial generators, Hilbert affected 
a change of paradigm that brought new methods. We present one of those methods here, 
and suggest more investigation. Our hope is that the historical drama, taken alongside 
the content of this article, will give physicists vital motivation to seriously 
consider new applications of XIX century techniques. 

\section*{Acknowledgments}
The author would like to thank Bill Harter for helpful discussions regarding 
molecular physics and representation theory. As well Neil J. A. Sloane for suggesting Molien sequences 
as a research topic, and the editors of the OEIS. This work is supported in part by a doctoral 
fellowship awarded by the University of Arkansas. 

\bibliographystyle{unsrt}
\bibliography{Correlation}

\end{document}